\documentclass[journal]{IEEEtran}

\usepackage{booktabs}
\usepackage{array}
\newcolumntype{L}[1]{>{\raggedright\arraybackslash}p{#1}}
\newcolumntype{C}[1]{>{\centering\arraybackslash}p{#1}}
\usepackage{graphicx}
\usepackage{url}
\usepackage[table]{xcolor}
\usepackage[hidelinks]{hyperref}
\begin{document}

\title{Energy as the Fourth NTN Service: Powering Night-Side HAPS from Orbit}

\author{Mukhtiar~Ahmad,
        and~Mohamed-Slim~Alouini,~\IEEEmembership{Fellow,~IEEE}%
\thanks{The authors are with the Computer, Electrical and Mathematical Science and Engineering Division, King Abdullah University of Science and Technology (KAUST), Thuwal, Saudi Arabia. E-mail: mukhtiar.ahmad@kaust.edu.sa.}}

\maketitle

\begin{abstract}
High-altitude platform stations (HAPS) promise persistent communication, navigation, and sensing (CNS) from the stratosphere. Their endurance is limited by the battery they carry through a 10--14~h night. It caps payloads at a few kilograms and confines operation to low latitudes and mild seasons. We argue that the night need not be powered entirely onboard. Low Earth orbit (LEO) satellites are sunlit for about two thirds of each orbit and routinely discard surplus solar energy. We propose an architecture in which LEO satellites act as orbital power banks and beam this surplus to night-side HAPS over near-infrared laser links, in direct mode when sunlit and in buffered mode when eclipsed. A control plane hands each HAPS across the tens of satellites in view and is designed for joint power/data scheduling. The architecture is forward-looking, and we state the evidence behind each part. Kilometer-scale optical power beaming and microradian pointing have been demonstrated. The link budget and the battery-mass saving are first-order calculated: in the nominal case the saving is 2.7~kg before reserve and 1.8~kg after reserve, and in the pessimistic case it is a loss. The contact geometry and the delivered energy are simulated. Kilowatt-class delivery from orbit, power-beam handover, the dual-band receiver, and certification remain open. Our three contributions are the orbital power-bank concept, a spot-matched dual-band receiver architecture, and a joint power/data scheduling mechanism.
\end{abstract}


\section{Introduction}
\label{sec:intro}

Non-terrestrial networks (NTNs) are moving to the center of cellular standardization. 3GPP has integrated satellite access into 5G, and low Earth orbit (LEO) mega-constellations serve millions of subscribers. High-altitude platform stations (HAPS), solar-powered aircraft and airships at about 20~km, form a middle tier with near-terrestrial latency and wide-area coverage \cite{haps-survey}. Both tiers can host communication, navigation, and sensing (CNS) services. A HAPS over a fixed region suits the most demanding of them: mission-critical services that must stay available when terrestrial infrastructure is congested, damaged, or absent.

However, this persistent stratospheric layer does not yet exist. The obstacle is not aeronautics or radio design but night-time energy. A solar HAPS must store by day enough energy to fly a 10--14~h night. The batteries claim a large share of the airframe mass and cap payloads at a few kilograms \cite{zephyr-specs} and a few hundred watts. The constraint compounds: more battery needs more wing, structure, and propulsion, hence more battery \cite{haps-flight-mech}. We call it the night-time energy bottleneck.

In this article we ask whether the night must be powered entirely onboard. LEO satellites run on a complementary schedule: a satellite at 550~km orbits about every 95~min and is sunlit for about 60~min of each orbit. Its arrays are overprovisioned and shunt the excess once the battery is full \cite{patel-power}. Long-range wireless power transfer (WPT) has also matured. Above 20~km the air is thin, dry, and nearly cloud-free, so a LEO-to-HAPS beam crosses almost none of the atmosphere that constrains ground beaming.

We present an architecture that delivers orbital energy to HAPS. To our knowledge, it is the first to treat that energy as a scheduled network resource: a fourth service beside communication, navigation, and sensing, so that CNS becomes CNS plus energy (CNS+E). Section~\ref{sec:readiness} states which of its elements are demonstrated, first-order calculated, simulated, or open. In the nominal case, one beam drawing 2~kW from the satellite array delivers 1.9~kWh across a 12~h night, 35~percent of a 450~W load, and costs the serving satellites 21~kWh. The battery-mass saving is 2.7~kg before reserve and 1.8~kg after reserve; it is a loss in the pessimistic case and 8.8~kg after reserve in the optimistic case (Section~\ref{sec:feasibility}).

Our three principal contributions are as follows. The first is the \emph{orbital power-bank concept}: LEO satellites beam live array surplus to a HAPS in direct mode when sunlit and discharge an onboard buffer in buffered mode when eclipsed. The second is a \emph{spot-matched dual-band receiver architecture}: one region of the wing is solar skin by day and a 1070~nm laser collector by night. The third is a \emph{joint power/data scheduling mechanism}: a control plane hands each HAPS across the satellites in view, shares satellites among HAPS, selects among orbital, ground, and stored sources, and divides each beam between power and backhaul data. The third is what turns power beaming into a network service.

\section{Night-Time Energy Bottleneck}
\label{sec:wall}

\subsection{Energy Constraint}

Consider a representative solar HAPS: a fixed-wing aircraft of about 25~m span and 65~kg airframe mass, cruising near 20~km \cite{zephyr-specs}. A 400--500~W load through a 12~h night needs about 5--6~kWh of usable storage. Even with 450~Wh/kg silicon-anode cells \cite{amprius}, depth-of-discharge, cold derating, and end-of-life margins leave about 370~Wh/kg usable. The battery then weighs about 15~kg before packaging. What remains for the CNS payload is about 5~kg \cite{zephyr-specs} and a few hundred watts.

\subsection{Constraint Severity}

Two properties make the constraint severe. First, it is self-amplifying: small increases in storage cause disproportionate airframe growth \cite{haps-flight-mech}, so a 20~percent gain in cell specific energy does not buy a 20~percent larger payload. Second, it is geographic and seasonal: in winter at 45--60$^\circ$ latitude the night runs past 14~h and the solar harvest shrinks. This confines persistent operation to a band of about $\pm$40$^\circ$. External energy at night widens that envelope.

\subsection{Onboard Alternatives}

Onboard alternatives exist (Table~\ref{tab:options}). Hydrogen fuel cells add storage and consumable logistics, airships trade this mass growth for station-keeping power, and better cells improve slowly. All of them supply the night from onboard resources; our architecture does not.

\begin{table*}[t]
\caption{Options for meeting the HAPS night-time energy requirement.}
\label{tab:options}
\centering
\footnotesize
\setlength{\tabcolsep}{3pt}
\begin{tabular}{@{}L{2.4cm}L{3.0cm}L{3.5cm}L{2.9cm}L{2.6cm}L{2.5cm}@{}}
\toprule
\textbf{Energy source} & \textbf{Night coverage} & \textbf{Added airframe mass} & \textbf{Weather and path dependence} & \textbf{Geographic envelope} & \textbf{Maturity} \\
\midrule
Onboard battery (baseline) & Full, if sized for the full night & High (about 15~kg, a quarter of the airframe) & None & Low latitude, mild seasons & Flight-demonstrated (limited) \\
\rowcolor{gray!12}
H$_2$ fuel cell & Full & Medium, plus pressurized or cryogenic storage & None & Wider, but consumable logistics & Laboratory and early flight \\
Ground-to-HAPS laser & Weather-limited & Low (receiver only) & Full troposphere; station beneath & Clear-sky sites only & Prototypes \\
\rowcolor{gray!12}
\textbf{Orbit-to-HAPS laser (this work)} & \textbf{Continuous (needs power-beam handover)} & \textbf{Low (2--3.5~kg of added hardware)} & \textbf{Negligible (above the troposphere)} & \textbf{Global, including high-latitude winter} & \textbf{Components shown; system open} \\
\bottomrule
\end{tabular}
\end{table*}

\section{Orbital Energy Delivery}
\label{sec:case}

\subsection{Complementary Energy Cycles}

LEO satellites and HAPS keep opposite schedules. A satellite at 550~km is sunlit for about 63~percent of each 95~min orbit and eclipsed for up to 35~min, 15~times a day. A sunlit satellite can deliver its surplus as it is generated. An eclipsed satellite must deliver from a buffer and pays a storage penalty. Section~\ref{sec:design} calls these two cases direct mode and buffered mode.

The source is also large: more than 10,000 Starlink satellites are already in orbit \cite{starlink-stats}. We assume a 5~kW array per satellite. From 20~km, tens of satellites sit above a 25$^\circ$ elevation mask at any time, so continuous delivery requires power-beam handover, the transfer of the beam from one satellite to the next.

\subsection{Technology Readiness}
\label{sec:readiness}

Three developments support this design. First, long-range optical power beaming has reached relevant scale. DARPA's POWER program sent more than 800~W over 8.6~km in 2025, and Star Catcher has since delivered more than 1.1~kW to standard solar panels in ground tests \cite{darpa-power, starcatcher}. Second, WPT to moving targets works. Xidian University's Zhuri project delivered 143~W to a drone in flight at 30~m, at microwave frequencies \cite{zhuri}. Third, the satellite side is mature where it matters. Microradian pointing is standard on optical inter-satellite links (ISLs) \cite{isl-pointing}, fiber lasers are an industrial technology, and Caltech's MAPLE experiment demonstrated WPT in orbit \cite{maple}. However, flown ISL terminals have apertures near 13~cm and milliwatt-to-watt output, so a 30~cm aperture and kilowatt-class output are the scale-ups required.

Table~\ref{tab:evidence} places every element of the architecture in one of four tiers. \emph{Demonstrated} means shown in hardware, at the stated scale. \emph{First-order calculated} means computed from published component figures, without simulation. \emph{Simulated} means obtained from our orbital and scheduling simulations (Section~\ref{sec:feasibility}). \emph{Open} means that neither a demonstration nor a validated model exists. A demonstrated element can still need scale-up, so the last column states what remains to be shown.

\begin{table*}[t]
\caption{Evidence status of the elements of the architecture.}
\label{tab:evidence}
\centering
\scriptsize
\setlength{\tabcolsep}{3.5pt}
\renewcommand{\arraystretch}{0.95}
\begin{tabular}{@{}L{3.3cm}L{3.5cm}L{5.1cm}L{5.4cm}@{}}
\toprule
\textbf{Element} & \textbf{Status} & \textbf{Evidence or basis} & \textbf{What remains to be shown} \\
\midrule
Long-range power beaming & Demonstrated (optical: ground; microwave: in orbit) & POWER: $>$800~W over 8.6~km \cite{darpa-power}; Star Catcher: $>$1.1~kW \cite{starcatcher}; MAPLE: mW-class, in orbit \cite{maple} & kW-class optical output through a 30~cm space aperture over 530--1,100~km \\
\rowcolor{gray!12}
Delivery to a moving receiver & Demonstrated (microwave, 30~m) & Zhuri: 143~W to a drone in flight \cite{zhuri} & optical carrier; hundreds of km; varying wing attitude \\
Microradian pointing from LEO & Demonstrated (data terminals) & optical ISLs \cite{isl-pointing}; apertures of about 13~cm & power-class beam; tracking on the HAPS side \\
\rowcolor{gray!12}
Link budget; battery-mass saving & First-order calculated & Table~\ref{tab:link}: published component figures, three sensitivity cases & measured dual-band cell; reserve sized from dependability data \\
Contact geometry; direct-mode window & Simulated & contact-geometry simulation (Section~\ref{sec:contact}) & validation against real ephemerides \\
\rowcolor{gray!12}
Delivered energy; number of handovers & Simulated & scheduler-in-the-loop simulation (Section~\ref{sec:contact}) & contention beyond a cluster; traffic competing with power; surplus on satellites just out of eclipse \\
Power-beam handover & Open & routine for ISL data & demonstration with a power beam \\
\rowcolor{gray!12}
Buffered-mode discharge & Open & high-rate cells exist & flight-qualified discharge at 2~kW \\
Dual-band receiver & Open & 55\% laser-tuned \cite{laser-pv} and 25\% broadband cells & balanced dual-band cell; compact region 3.5~m across \\
\rowcolor{gray!12}
Safety, security, certification & Open & laser-clearinghouse and ISL practice & mechanisms of Section~\ref{sec:agenda}; no regulatory pathway \\
\bottomrule
\end{tabular}
\end{table*}

\subsection{Carrier Selection}

Carrier choice follows from diffraction: the spot grows with wavelength and range and shrinks with aperture. At 5.8~GHz over 500~km, even a 10~m aperture gives a 2.6~km spot, and a few square meters of rectenna capture about one part in a million. Microwave WPT thus suits short ranges \cite{zhuri} or kilometer-scale apertures \cite{maple}. Even at millimeter-wave frequencies with a 100~m aperture, a wing-scale rectenna of the kind proposed for aircraft \cite{ssp-aea} captures about 1~percent. At 1070~nm, a beam from a 30~cm telescope forms a 3--7~m spot, matched to a wing receiver, and laser-tuned photovoltaic (PV) converters reach about 55~percent efficiency \cite{laser-pv}. We therefore assume near-infrared laser delivery.

\section{System Architecture}
\label{sec:design}

\begin{figure}[t]
  \centering
  \includegraphics[width=\linewidth]{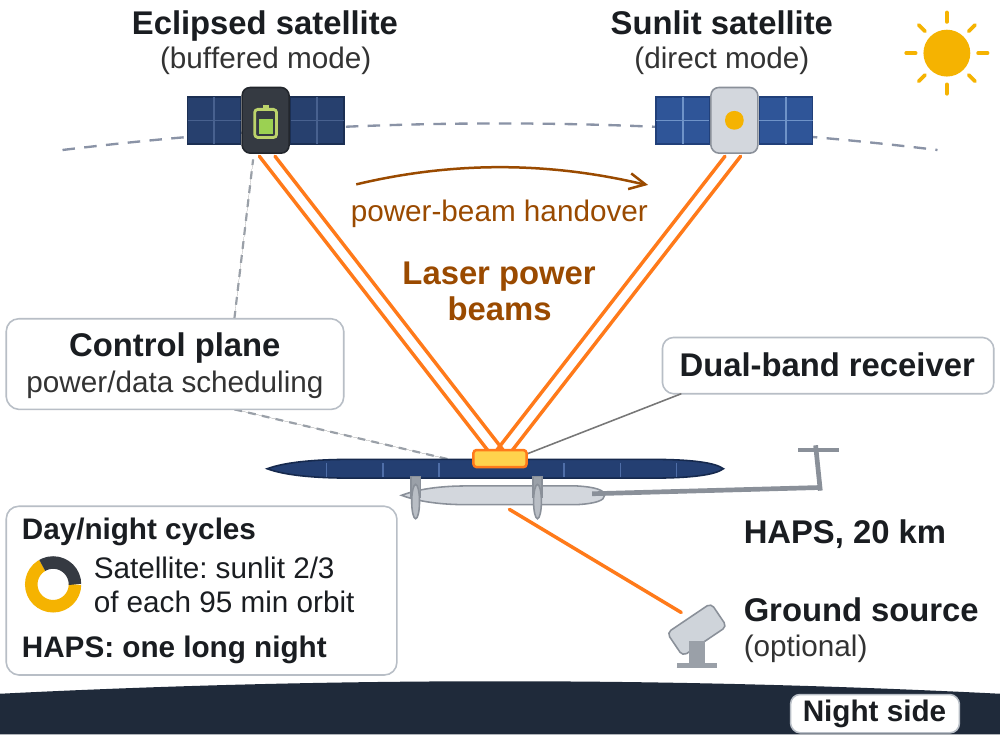}
  \caption{Proposed architecture. An orbital power bank beams in direct mode when sunlit and in buffered mode when eclipsed. Power-beam handover passes the HAPS from one power bank to the next; one power bank serves at a time. The dual-band receiver converts sunlight by day and the 1070~nm beam by night. Inset: day/night cycles.}
  \label{fig:arch}
\end{figure}

Fig.~\ref{fig:arch} sketches the architecture: orbital power banks, a spot-matched dual-band receiver, and a control plane for joint power/data scheduling.

\subsection{Orbital Power Banks}
\label{sec:banks}

Geometry dictates two delivery modes. In \emph{direct mode}, a sunlit power bank beams live array surplus to a HAPS in view, with no storage and so no round-trip loss or cycle aging. Every satellite above the elevation mask is a \emph{candidate}. A satellite at 550~km stays sunlit until the local solar depression reaches about 23$^\circ$, so the direct-mode window outlasts local daylight. Deeper into the night every candidate is eclipsed, and \emph{buffered mode} applies: a power bank charges a buffer during sunlit arcs and discharges it through its laser on night-side passes. Each power bank carries a kWh-class buffer, a kW-class 1070~nm fiber laser, and a 30~cm gimbaled telescope that tracks a beacon on the HAPS.

\subsection{Receiver Design}
\label{sec:receiver}

The receiver's area follows from the beam. A 30~cm transmit aperture forms a spot about 3.4~m across at zenith, widening to 7.0~m at the longest slant ranges. A compact region of about 12~m$^2$, about 3.5~m across, captures 0.92 of the incident Gaussian beam at zenith; a 2.5~m wide panel loses about 7~percent of the delivered energy. A slender wing of about 1~m chord captures only 0.44 at zenith, and the delivered energy halves (Section~\ref{sec:sens}). A Zephyr-class airframe therefore needs a wider center panel.

Spot-matching brings three properties that do not depend on cell efficiency. The first is safety. At high elevation, under 10~percent of the beam misses the receiver and continues to the ground, at about 7~W/m$^2$. Near the 25$^\circ$ elevation mask, about three quarters of the beam misses. The elevation mask is therefore a safety constraint as much as an efficiency one. The second is pointing. The receiver is slightly larger than the spot at high elevation, so pointing error degrades capture gradually: a fixed offset of 1~$\mu$rad costs about 4~percent of capture and 2~$\mu$rad about 18~percent. The third is thermal: spread over 12~m$^2$, the absorbed heat stays well under the daytime solar flux the wing already tolerates (Section~\ref{sec:power}).

To keep the area generating by day, the spot-matched region uses \emph{dual-band} cells: one laminate that converts broadband sunlight by day and the 1070~nm line by night. The cost is PV conversion efficiency, since a dual-band cell cannot match a device dedicated to one band. Laser-tuned converters reach about 55~percent \cite{laser-pv} and broadband cells about 25~percent; a balanced dual-band cell is the main open component. The receiver normal also moves with aircraft attitude. Our evaluation covers this with a fixed allowance (Section~\ref{sec:power}).

\subsection{Joint Power/Data Scheduling and Source Selection}
\label{sec:scheduling}

The control plane is what makes this a networking system. A night-side HAPS sees tens of candidate power banks, each with its own illumination, buffer level, and geometry. Continuous supply is a sequence of serving sessions, each about 1~min long in our simulation, linked by power-beam handover. In each interval the scheduler chooses which power bank serves which HAPS, in which mode, and how to divide that power bank's beam between power and backhaul data. It must also leave each power bank enough buffer for its own next eclipse. We call this joint power/data scheduling. The split is the new degree of freedom: contact-plan design and beam placement assign links over time, but neither divides one link between a storable commodity (energy) and deadline-bound traffic (data). Carrying both on one beam costs little: giving 5~percent of the beam to a co-aligned data carrier costs about 8~W of the 160~W night-average delivered power. The data share must also meet the requirements of the transport network it feeds, so joint power/data scheduling needs transport-aware design as well as link-level scheduling.

Our evaluation covers a narrower scope than this mechanism. Section~\ref{sec:feasibility} evaluates energy-delivery scheduling only: it decides which power bank serves the HAPS and when to hand over, and it holds the data share fixed at 5~percent. Jointly optimizing energy delivery and communication traffic, with traffic deadlines and source selection in the loop, is the general problem and remains future work (Section~\ref{sec:agenda}).

The mechanism also selects among sources. A patch on the belly of the HAPS admits an opportunistic ground uplink from a laser beneath it. This link is weather-limited and tied to fixed sites, so it cannot anchor the night budget. However, ground-to-stratosphere beaming has been analyzed in detail \cite{rand-tr898} and gives an early deployment path. Power banks can also feed one another by inter-satellite beaming \cite{starcatcher}.

\subsection{Deployment Paths}

Deployment can proceed in three stages: terminals and modest buffers hosted on a communication constellation; purpose-built power banks phased for night-side contact over HAPS corridors; and inter-satellite energy exchange across fleets. The first stage still needs the power-class pointing and power-beam handover of Table~\ref{tab:evidence}.

\section{Feasibility Analysis}
\label{sec:feasibility}

We present a first-order analysis, with the contact geometry and the delivered energy obtained by simulation. A beam is specified by the power it draws from the satellite array: a 2~kW beam carries about 0.85~kW of optical power to the wing. Table~\ref{tab:link} gives the link budget in pessimistic, nominal, and optimistic cases and carries each through to delivered energy, satellite-side energy, and battery mass.

\begin{table}[t]
\caption{Link budget and sensitivity for one beam drawing 2~kW from an 8,000-satellite shell (12~h equinox night at 35$^\circ$N, 450~W load).}
\label{tab:link}
\centering
\footnotesize
\setlength{\tabcolsep}{3pt}
\renewcommand{\arraystretch}{1.1}
\begin{tabular}{@{}L{3.55cm}C{0.66cm}C{0.66cm}C{0.66cm}L{2.5cm}@{}}
\toprule
\textbf{Parameter} & \textbf{Pess.} & \textbf{Nom.} & \textbf{Opt.} & \textbf{Basis} \\
\midrule
\multicolumn{5}{@{}l}{\textit{Stage efficiencies}} \\
Laser efficiency & 0.35 & 0.45 & 0.50 & fiber laser \\
Path transmittance & 0.93 & 0.95 & 0.97 & above 20~km \\
Geometric capture* & \multicolumn{3}{c}{0.86} & over served passes \\
Pointing and incidence & 0.80 & 0.90 & 1.00 & 1~$\mu$rad jitter (nom.) \\
PV conversion efficiency & 0.25 & 0.30 & 0.45 & dual-band (assumed) \\
Buffer round trip & 0.85 & 0.90 & 0.95 & buffered mode only \\
\textbf{End-to-end efficiency*} & \textbf{0.05} & \textbf{0.09} & \textbf{0.18} & night average \\
\midrule
\multicolumn{5}{@{}l}{\textit{Serving time}} \\
Handover dead time (\%)* & 28 & 14 & 0 & 20/10/0~s each \\
\midrule
\multicolumn{5}{@{}l}{\textit{Outputs}} \\
Delivered energy (kWh) & 0.9 & 1.9 & 4.3 & 16/35/80\% of load \\
Satellite-side energy (kWh) & 17 & 21 & 24 & 2~kW while serving \\
Cells displaced (kg) & 2.4 & 5.2 & 11.8 & 370~Wh/kg usable \\
Added hardware (kg) & 3.5 & 2.5 & 2.0 & subtracted \\
\textbf{Saving before reserve (kg)} & \textbf{$-$1.1} & \textbf{2.7} & \textbf{9.8} & \\
Onboard reserve (kg) & 0.6 & 0.9 & 1.0 & 3/2/1~h; subtracted \\
\textbf{Net saving after reserve (kg)} & \textbf{$-$1.7} & \textbf{1.8} & \textbf{8.8} & \\
\bottomrule
\end{tabular}

\vspace{2pt}
\parbox{\linewidth}{\footnotesize\raggedright *~From the scheduler-in-the-loop simulation; the other inputs are first-order. Pess., Nom., Opt.: pessimistic, nominal, and optimistic cases. Outputs are computed from unrounded values.}
\end{table}

\subsection{End-to-End Efficiency and Delivered Power}
\label{sec:power}

The end-to-end efficiency in Table~\ref{tab:link} is the product of the stage efficiencies listed above it. In the nominal case, laser efficiency (0.45), path transmittance (0.95), geometric capture (0.86), the pointing-and-incidence factor (0.90), and PV conversion efficiency (0.30) multiply to 0.10. Geometric capture is the share of the beam that lands on the receiver. It is 0.92 at zenith and 0.22 at the elevation mask, and 0.86 is its average over the passes the scheduler actually serves. The pointing-and-incidence factor removes what pointing jitter and wing attitude cost: root-mean-square (RMS) jitter of about 1~$\mu$rad, which operational ISLs meet \cite{isl-pointing}, costs about 8~percent, and a 10$^\circ$ wing bank costs under 2~percent. PV conversion efficiency sets how much of the captured light becomes electricity. In buffered mode, which covers two thirds of the serving time, the buffer round trip (0.90) also applies, so the night-average end-to-end efficiency is 0.09.

Handover dead time is the interval in which the beam is off while the next power bank acquires the receiver. It does not change the end-to-end efficiency; it shortens the serving time. With a dead time of 14~percent, the serving power banks draw 21~kWh over the night instead of 24~kWh, and 9~percent of that, 1.9~kWh, reaches the HAPS. A high-elevation pass delivers about 210~W, and the average over the 12~h night, including dead time, is about 160~W.

The 0.85~kW of optical power gives about 65~W/m$^2$ averaged over the receiver and 190~W/m$^2$ at the spot center, or 0.07--0.2~sun. The absorbed heat averages 46~W/m$^2$ for one beam and 92~W/m$^2$ for two, with peaks near 130 and 260~W/m$^2$.

\subsection{HAPS-Side Benefits}
\label{sec:benefits}

The nominal 1.9~kWh a night displaces 5.2~kg of cells at the usable 370~Wh/kg of Section~\ref{sec:wall}. Two masses are subtracted from this figure. The first is added hardware: about 1~kg for the belly patch and power conditioning plus 1--2.5~kg of thermal spreading, 2.5~kg in total in the nominal case. The dual-band cells replace laminate the wing already carries. Subtracting the hardware gives the \emph{saving before reserve}, 2.7~kg in the nominal case. The second is an onboard reserve sized to the longest credible outage, because storage can be shed only as far as delivery is dependable. A 2~h reserve of the delivered power (0.3~kWh, 0.9~kg) leaves the \emph{net saving after reserve}, 1.8~kg in the nominal case.

The sensitivity columns of Table~\ref{tab:link} show how much this depends on the receiver. In the optimistic case the saving is 9.8~kg before reserve and 8.8~kg after reserve. In the pessimistic case it is a loss of 1.1~kg before reserve and 1.7~kg after reserve. A degraded receiver alone (pointing-and-incidence factor 0.80 and PV conversion efficiency 0.25) delivers 1.4~kWh and saves 1.3~kg before reserve and 0.7~kg after reserve. In these two cases the service does not pay for itself in mass, or barely does. Its value is then the 70--120~W of night-time payload power it supplies without storage, and the wider latitude and season envelope that power opens.

\subsection{Contact Geometry, Handover, and Delivered Energy}
\label{sec:contact}

We now make the analysis concrete for a HAPS served by power banks on a 550~km shell with a 25$^\circ$ elevation mask. Analytically, a satellite is usable within a cap that covers 0.51~percent of the shell, so a uniformly spread shell of 8,000 satellites would place about 40 in view.

\begin{figure}[t]
\centering
\includegraphics[width=\linewidth]{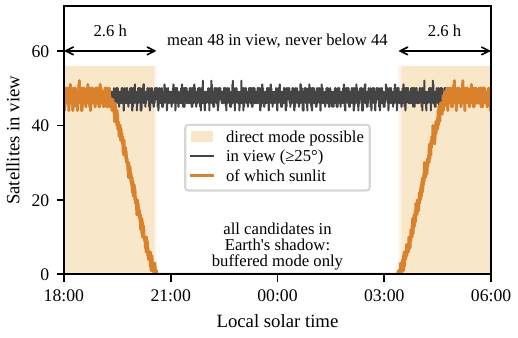}
\caption{Contact-geometry simulation (no link model): satellites above the 25$^\circ$ elevation mask (gray) and the sunlit subset (orange) through a 12~h equinox night at 35$^\circ$N, for an 8,000-satellite Walker shell at 550~km and 53$^\circ$. Direct mode is possible while the orange curve is above zero, about 2.6~h at each edge of the night. In between, only buffered mode is possible.}
\label{fig:sim}
\end{figure}

A contact-geometry simulation replaces the uniform assumption with orbits (Fig.~\ref{fig:sim}): 8,000 satellites in 100 planes of 80 at 53$^\circ$ inclination, on Keplerian circular orbits with a cylindrical Earth shadow, propagated in 10~s steps through a 12~h equinox night over a HAPS at 35$^\circ$N. It gives three results. The first is the number of candidates: a mean of 48 and never fewer than 44 (gray curve). The second is the number of sunlit candidates, which can serve in direct mode (orange curve): all of them for the first and last 1.25~h of the night, at least one for about the first and last 2.6~h (5.1~h in total), and none in the middle 6.9~h. The fall of the orange curve to zero is satellite eclipse, not a loss of satellites. The third is the number of candidate passes: 9,737 cross the night, one for every time a satellite rises above the mask, and they last 3.5~min on average and 4.7~min at most. Candidate passes are opportunities, not handovers. About 48 are open at any instant, and the HAPS is served by one power bank at a time.

\begin{figure}[t]
  \centering
  \includegraphics[width=\linewidth]{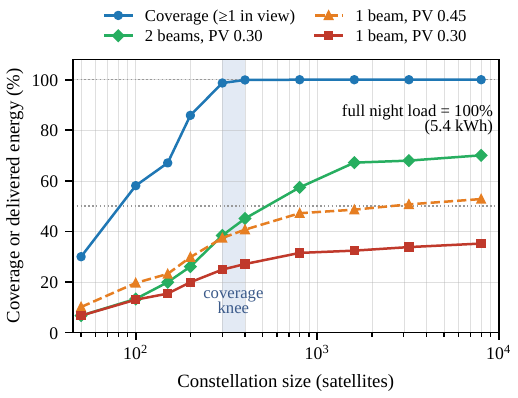}
  \caption{Scheduler-in-the-loop simulation: coverage (share of the night with at least one candidate in view) and delivered energy (share of the 5.4~kWh night load) versus constellation size, for one HAPS at 35$^\circ$N. Each beam draws 2~kW. PV 0.30 is the nominal PV conversion efficiency of Table~\ref{tab:link} and PV 0.45 the optimistic one. The two-beam curve is at PV 0.30. Handover dead time is included.}
  \label{fig:sweep}
\end{figure}

A scheduler-in-the-loop simulation adds the link model and quantifies the delivered energy (Fig.~\ref{fig:sweep}, Table~\ref{tab:link}). The constellation is a Walker-delta shell at 550~km and 53$^\circ$ with generic phasing (phasing factor 1, not tuned to a corridor). The HAPS, mask, night, and time step are those of Fig.~\ref{fig:sim}. In each step the scheduler keeps the serving power bank while it stays above the mask and no other candidate offers 30~percent more delivered power. Otherwise it hands over to the best candidate. Delivered power follows from the capture model of Section~\ref{sec:receiver} and the nominal link budget of Table~\ref{tab:link}. This rule favors high elevation (70$^\circ$ on average) and sunlit satellites. Every sunlit satellite is assumed to have 2~kW of surplus, which satellites just out of eclipse may not. The buffer is never binding, since each power bank serves for about 1~min. There is no contention beyond the co-located cluster below, no priority class, and no weather (the ground uplink is not simulated). Backhaul traffic does not compete with power: the 5~percent data share is applied analytically.

A handover is one change of serving power bank. The scheduler makes about 610 handovers a night, so it uses about 610 of the 9,737 candidate passes, or 6~percent, each from a different satellite. A serving session lasts about 1~min, shorter than the 3.5~min pass, because the scheduler hands over as soon as another candidate offers 30~percent more power. Each handover turns the beam off for 10~s in the nominal case (20~s in the pessimistic case and 0~s in the optimistic case, with make-before-break handover). The 610 handovers therefore interrupt delivery for 1.7~h in total, which is the 14~percent handover dead time of Table~\ref{tab:link}. A rule that re-selects the best candidate in every step would make about 1,700 handovers and lose 39~percent of the night. The scheduler serves about 3.4~h of the night in direct mode and the rest in buffered mode.

The burden on each satellite is light in energy and heavy in power. The serving power banks together draw about 21~kWh per night for one HAPS, or about 34~Wh for each of the 610 satellites that serve. This is well under 1~percent of the 50--100~kWh that a 5~kW array generates per day. The demanding part is discharge rate: 2~kW from a 1--2~kWh buffer needs high-rate cells and spends cycles.

Fig.~\ref{fig:sweep} shows the result against constellation size. Continuous coverage of one HAPS emerges near 300--400 satellites. Beyond that knee, one 2~kW beam delivers 1.5--1.9~kWh against a 5.4~kWh night load (450~W for 12~h), or 27--35~percent. The beam is power-limited, not coverage-limited: the full night needs about three times the power. Two beams from two power banks onto the same receiver reach 45--70~percent, and one beam at the optimistic PV conversion efficiency of 0.45 reaches 41--53~percent. For a co-located cluster, with each HAPS holding a different power bank, the 8,000-satellite shell serves 20~HAPS at 1.6~kWh each and 44 at 1.2~kWh. Availability holds at 100~percent up to 44~HAPS, the minimum number in view. A dedicated 400-satellite shell serves about one.

\subsection{Sensitivity and Alternative Scenarios}
\label{sec:sens}

Between the pessimistic and optimistic columns of Table~\ref{tab:link}, the delivered energy spans 0.9--4.3~kWh and the net saving after reserve spans $-$1.7 to $+$8.8~kg. The satellite-side energy moves far less, 17--24~kWh, since the array draw is fixed. The end-to-end efficiency therefore sets the benefit to the HAPS, not the cost to the constellation.

Two alternative scenarios test the nominal case. The first is a winter-solstice night at 45$^\circ$N, which lasts 15.4~h with a load of 6.9~kWh. The same beam delivers 2.4~kWh, again 35~percent, with 4.2~h in direct mode. The second is a slender wing, the harder case. With the same beam on a 1~m by 12~m strip, geometric capture at zenith falls from 0.92 to 0.44 and nominal delivery falls to 0.9~kWh, 17~percent of the night load. The saving becomes a loss of 0.1~kg before reserve and 0.5~kg after reserve. A 2.5~m by 4.8~m panel recovers 1.8~kWh.

\section{Open Problems and Research Directions}
\label{sec:agenda}

The elements marked open in Table~\ref{tab:evidence}, and the limits of our evaluation, define four research themes. They are future work, not supported findings.

\textbf{Hardware, pointing, and power-beam handover.} Two hardware items are open: kilowatt-class output through a 30~cm space aperture, and a balanced dual-band cell. Both ends of the link move. The satellite travels at 7.6~km/s and needs point-ahead and ephemeris feed-forward. The HAPS faces winds and attitude changes that tilt and sometimes shadow the receiver. No combined pointing budget exists for power-class beams. Power-beam handover is routine for ISL data but undemonstrated for power.

\textbf{Joint power/data scheduling and energy management.} The control plane must decide online which power bank serves which HAPS, with admission control, fairness, and preemption under contention. Two problems go beyond our evaluation. The first is joint power/data scheduling with traffic in the loop: energy is storable and data has deadlines, yet both share apertures and contact windows. The second is battery health: buffered-mode discharge ages a power bank's cells, so a schedule that is best for each pass can be poor over a lifetime.

\textbf{Safety, security, and deconfliction.} Energy delivery opens attack surfaces that data links do not have, because the beam itself is the hazard. A spoofed beacon could lure it to an unauthorized receiver, jamming the beacon denies charging, and a forged schedule or ephemeris is a forged beam. We see six mechanisms as the minimum. First, mutual authentication: a challenge-response exchange on a spread-spectrum beacon, keyed to the serving session, so that power bank and receiver each accept only the expected peer. Second, beam authorization: the laser fires only with a token signed by the control plane that names the power bank, the HAPS, the time window, the elevation mask, and the maximum power. The ephemerides and schedules behind it are signed and access-controlled. Third, fail-safe shuttering: loss of beacon track, a receiver fault, or a corridor intrusion closes the shutter within one control cycle, and the shutter state is reported in signed telemetry. Fourth, exclusion-zone management: the part of the beam that misses the receiver reaches the ground up to tens of kilometers away. Every serving session is therefore screened against aircraft and spacecraft catalogs with predictive shuttering, as in laser-clearinghouse practice, and this footprint sets the elevation mask. Fifth, independent tracking verification: a second sensor on the HAPS checks the beam position that the fine-steering loop reports, and disagreement closes the shutter. Sixth, tamper-evident metering and audit logging: the token, pointing record, delivered energy, and shutter events of every serving session are logged and signed. Coexistence with optical communication terminals and astronomy adds similar exclusion constraints.

\textbf{Regulation and economics.} No certification pathway exists today. We expect one to combine three regimes: laser safety standards (IEC 60825-1, ANSI Z136) for the beam and its footprint, airspace coordination with civil aviation authorities, and ITU-R spectrum and orbital coordination for the data carrier and the constellation. Multi-operator deployment needs standard interfaces and settlement for delivered energy, based on the audit log. Whether launch-amortized orbital energy beats marginal battery mass needs a techno-economic model.

\section{Related Work}
\label{sec:related}

We organize prior work by where the energy travels. Orbit-to-ground space solar power has been studied for decades; Caltech's MAPLE experiment validated lightweight transmit arrays in orbit \cite{maple}. Ground-to-stratosphere delivery was analyzed at RAND \cite{rand-tr898} and is pursued by DARPA's POWER program \cite{darpa-power}, but such links cross the full troposphere and need a station beneath the HAPS. Orbit-to-orbit delivery is pursued by Star Catcher \cite{starcatcher}. Within the stratosphere, the Stratospheric Grid proposal networks energy among the HAPS themselves \cite{stratgrid}; ours supplies that layer from above. Closest is orbit-to-aircraft delivery: Goh \emph{et al.} propose microwave delivery from a LEO constellation to all-electric aircraft \cite{ssp-aea}. We differ in target (a solar HAPS drawing hundreds of watts), in carrier (a laser with a spot of a few meters), and in contribution (network-level joint power/data scheduling, not link-level beam tracking).

\section{Conclusions}
\label{sec:conclusion}

We have argued that the night-time energy bottleneck follows from supplying the night from onboard storage, and that it relaxes when orbital solar surplus is delivered as a scheduled service. Our analysis supports three findings. In our simulation, a shell of more than a few hundred satellites keeps a candidate power bank in view of the HAPS all night. In the nominal case, one beam drawing 2~kW supplies 35~percent of the night load, and two beams about 70~percent. The battery-mass saving is modest, 2.7~kg before reserve and 1.8~kg after reserve, and it depends on the receiver. The rest is future work: kilowatt-class delivery from orbit, power-beam handover, the dual-band receiver, certification, and joint power/data scheduling with traffic in the loop. Of these, the control plane falls to the networking community: it must schedule energy as a fourth satellite service, with the rigor now reserved for spectrum and capacity.

\section*{Acknowledgment}
The authors used the AI assistant Claude (Anthropic) for language editing to improve clarity and readability. They reviewed all content and take full responsibility for it.

\begin{IEEEbiographynophoto}{Mukhtiar Ahmad}
is a Research Scientist with the Computer, Electrical and Mathematical Science and Engineering Division, King Abdullah University of Science and Technology (KAUST), Thuwal, Saudi Arabia. His research interests include cellular networks, distributed systems, non-terrestrial networks, and mission-critical and ultra-reliable low-latency communication.
\end{IEEEbiographynophoto}

\begin{IEEEbiographynophoto}{Mohamed-Slim Alouini}
(Fellow, IEEE) received the Ph.D. degree in electrical engineering from the California Institute of Technology in 1998. He is a Distinguished Professor of Electrical and Computer Engineering at KAUST, Thuwal, Saudi Arabia. His research interests include the modeling, design, and performance analysis of wireless communication systems.
\end{IEEEbiographynophoto}

\end{document}